\documentclass{icrc2009}

\usepackage{graphicx}   % for including figures
\usepackage[caption=false]{caption}    % for captions
\usepackage[font=footnotesize]{subfig} % subfig.sty for a double column floating figure using two subfigures
\usepackage{fixltx2e}
\usepackage{url}

\newcommand{\shorttitle}[1]%
{\markboth{Proceedings of the 31\MakeLowercase{$^{st}$} ICRC, {\L}\'{o}d\'{z} 2009}{#1} }
\newcommand{\etal}{\MakeLowercase{\textit{et al. }}} % "et al."

\begin{document}
\title{Observation of GRBs with AGILE}

\author{\IEEEauthorblockN{M.~Marisaldi\IEEEauthorrefmark{1}, 
                          G.~Barbiellini\IEEEauthorrefmark{2}\IEEEauthorrefmark{3}, 
                          E.~Costa\IEEEauthorrefmark{4}, 
                          S.~Cutini\IEEEauthorrefmark{5}, 
                          E.~Del~Monte\IEEEauthorrefmark{4}, 
                          I.~Donnarumma\IEEEauthorrefmark{4}\\
                          Y.~Evangelista\IEEEauthorrefmark{4},
                          M.~Feroci\IEEEauthorrefmark{4}, 
                          F.~Fuschino\IEEEauthorrefmark{1}, 
                          M.~Galli\IEEEauthorrefmark{6}, 
                          A.~Giuliani\IEEEauthorrefmark{7}, 
                          C.~Labanti\IEEEauthorrefmark{1}, 
                          I.~Lapshov\IEEEauthorrefmark{4}\IEEEauthorrefmark{8}\\
                          F.~Lazzarotto\IEEEauthorrefmark{4}, 
                          P.~Lipari\IEEEauthorrefmark{9}\IEEEauthorrefmark{10}, 
                          F.~Longo\IEEEauthorrefmark{2}, 
                          S.~Mereghetti\IEEEauthorrefmark{7}, 
                          E.~Moretti\IEEEauthorrefmark{2}, 
                          L.~Pacciani\IEEEauthorrefmark{4}, 
                          M.~Rapisarda\IEEEauthorrefmark{11}\\
                          P.~Soffitta\IEEEauthorrefmark{4},
                          M.~Tavani\IEEEauthorrefmark{4}\IEEEauthorrefmark{12}, 
                          M.~Trifoglio\IEEEauthorrefmark{1}, 
                          S.~Vercellone\IEEEauthorrefmark{13}\\ 
(on behalf of the AGILE Team)}

                            \\
\IEEEauthorblockA{\IEEEauthorrefmark{1} INAF-IASF Bologna, Via Gobetti 101, I-40129 Bologna, Italy}
\IEEEauthorblockA{\IEEEauthorrefmark{2} Dipartimento di Fisica Universit\`a di Trieste, via A. Valerio 2, I-34127 Trieste, Italy}
\IEEEauthorblockA{\IEEEauthorrefmark{3} INFN Trieste, via A. Valerio 2, I-34127 Trieste, Italy}
\IEEEauthorblockA{\IEEEauthorrefmark{4} INAF-IASF Roma, via del Fosso del Cavaliere 100, I-00133 Roma, Italy}
\IEEEauthorblockA{\IEEEauthorrefmark{5} ASI Science Data Center, Via E. Fermi 45, I-00044 Frascati (Roma), Italy}
\IEEEauthorblockA{\IEEEauthorrefmark{6} ENEA, via Martiri di Monte Sole 4, I-40129 Bologna, Italy}
\IEEEauthorblockA{\IEEEauthorrefmark{7} INAF-IASF Milano, via E. Bassini 15, I-20133 Milano, Italy}
\IEEEauthorblockA{\IEEEauthorrefmark{8} IKI, Moscow, Russia}
\IEEEauthorblockA{\IEEEauthorrefmark{9} INFN Roma ``La Sapienza'', p.le Aldo Moro 2, I-00185 Roma, Italy}
\IEEEauthorblockA{\IEEEauthorrefmark{10} Dipartimento di Fisica, Universit\`a La Sapienza, p.le Aldo Moro 2, I-00185 Roma, Italy}
\IEEEauthorblockA{\IEEEauthorrefmark{11} ENEA Frascati, via Enrico Fermi 45, I-00044 Frascati(Roma), Italy}
\IEEEauthorblockA{\IEEEauthorrefmark{12} Dipartimento di Fisica, Universit\`a Tor Vergata, via della Ricerca Scientifica 1, I-00133 Roma, Italy}
\IEEEauthorblockA{\IEEEauthorrefmark{13} INAF-IASF Palermo, Via Ugo La Malfa 153, 90146 Palermo, Italy}}

% please write the presenter's name and short title (3-4 words maximum)
%    which will appear at the header of the even pages.
\shorttitle{M. Marisaldi \etal GRBs with AGILE}
\maketitle

\begin{abstract}
Since its early phases of operation, the AGILE satellite is observing Gamma Ray Bursts (GRBs) over an energy range potentially spanning six orders of magnitude. In the hard X-ray band the SuperAGILE imager provides localization of about one GRB/month plus the detection of 1--2 GRBs per month out of its field of view. The Mini-Calorimeter detects about one GRB/week in the 350~keV--100~MeV energy range, plus several other transients at very short time scales. In fact, the on-board MCAL trigger logic, implemented for the first time on time windows as short as 300 microseconds, is particularly suitable for very short bursts detection. The Gamma-Ray Imaging Detector (GRID), sensitive in the 30~MeV--30~GeV range, firmly detected three GRBs (GRB~080514B, GRB~090401B and GRB~090510) plus some other candidates at a lower significance level. Moreover, all GRBs localized by other spacecrafts inside the GRID field of view are currently searched for possible detection, and upper limits are provided. In this paper we review the status of the observation of GRBs with AGILE, with particular emphasis on high energy and short time scales detections.
  \end{abstract}

 
\section{Introduction}
 
 The AGILE satellite \cite{Tavani2008}, an Italian space mission dedicated to high energy astrophysics launched on 23 April 2007, has the study of GRBs among its main scientific targets. 
The GRID (Gamma-ray Imaging Detector) instrument is the core of the AGILE mission; it is a pair conversion telescope composed of a tungsten-silicon tracker \cite{Prest2003} and a CsI(Tl) Mini-Calorimeter,  operating in the 30~MeV -- 30~GeV energy band, with a good sensitivity and angular resolution. Its large field of view ($60^\circ$x$60^\circ$) makes it a valuable instrument for GRB detection in the poorly explored gamma-ray energy band.
SuperAGILE \cite{Feroci2007} is the hard X-ray monitor of AGILE and is a twice 1-D coded
aperture instrument working in the 20 -- 60 keV energy band, with a
field of view of about 1~sr, an angular resolution of 3~arcmin and a
dead time of $5 \mu$s. 
The Mini-Calorimeter (MCAL) \cite{Labanti2009} is composed of 30 CsI(Tl) scintillator bars (dimensions: $\mathrm{15x23x375~mm^3}$ each) arranged in two orthogonal layers, for a total thickness of 1.5 radiation lengths. In a
bar the readout of the scintillation light is accomplished
by two custom PIN Photodiodes (PD) coupled one at each short side of the bar. In addition to being a subsystem of the GRID, MCAL is also equipped with a self-triggering operative mode and on-board logic making it an all-sky monitor in the 350~keV -- 100~MeV energy range.
The payload is surrounded and completed by a plastic anti-coincidence shield \cite{Perotti2006} for background charged particle rejection.

GRBs and other X-ray transients are a primary scientific goal of AGILE. For this reason the Payload Data Handling Unit contains
specific algorithms to trigger GRBs both in MCAL and in the SuperAGILE ratemeters.
A simultaneous GRB detection with GRID, MCAL and SuperAGILE would allow spectral coverage over six orders of magnitude.

\section{GRB detection with AGILE}

Table \ref{tab1} summarizes the main AGILE GRB detection results. Since the launch, more than 120~GRBs have been detected by either of the AGILE subsystems in different energy bands, the most remarkable being those detected by the GRID above several tens of MeV.

  \begin{table}[!h]
  \caption{Summary of AGILE GRB detections updated to 10 May 2009.}
  \label{tab1}
  \centering
  \begin{tabular}{lr}
  \hline
   AGILE detection & Number of events\\
   \hline 
   SuperAGILE localizations  &  21 \\
   MCAL detections           & 103  \\
   MCAL det. localized by SuperAGILE  &  2\\
   MCAL det. localized by Swift  & 21  \\
   MCAL det. localized by Fermi-GBM  &  15\\
   MCAL det. localized by INTEGRAL  &  1\\
   MCAL det. localized by IPN  &  10\\
  GRID detections           &   3  \\
  \hline
  \end{tabular}
  \end{table}

\subsection{GRB detection with SuperAGILE}

SuperAGILE is currently localizing about 1 GRB per month
and started in July 2007, with the GRB 070724B \cite{DelMonte2007} when AGILE was still in its Commissioning Phase. Up
to May 2009, SuperAGILE localized 21 GRBs, of which 15 were in the
twice 1-D area of the field of view and 6 in the external 1-D
region. Moreover, SuperAGILE detects about 1 -- 2 GRBs/month outside
the field of view and these events are provided to the  $3^{rd}$ Interplanetary
Network (IPN)\footnote{IPN web page: http://www.ssl.berkeley.edu/ipn3/}  to find the position using the triangulation technique.
 
\subsection{GRB detection with MCAL}
 
Between $22^{nd}$ June 2007 and $10^{th}$ May 2009 MCAL detected 103~GRBs, with an average detection rate of about 1~GRB/week. Most of these detections have been independently confirmed by other instruments. 59 detected events have been localized, either by Swift, SuperAGILE or the IPN, as reported in Table \ref{tab1} (in this table a GRB is assigned to the spacecraft that announced it first; some GRBs have been localized by more than one spacecraft). As SuperAGILE, also MCAL joined the IPN. The IPN localizations reported here are those publicly available in GCN circulars at the time of writing; since most of the MCAL events have also been detected by other IPN instruments the number of IPN localizations is expected to rise when the IPN catalogues become available, and can be obtained for bursts of particular interest. 

Due to programmatic constraints it was not possible to switch on and configure the on-board trigger logic prior to the end of November 2007. Then it was switched off again during January 2008, and it is steadily active since the $5^{th}$ of February 2008. When the trigger logic was not active, GRBs were detected by on-ground analysis, scanning the scientific ratemeters data (11~energy bands spectra accumulated every 1.024~seconds)  for rate increases with a dedicated software task. Despite several GRBs having been detected with this method, the coarse time and energy binning limits the scientific exploitation of the data. On the contrary, with the onset of the on-board trigger logic, data are downloaded in photon-by-photon mode after a valid trigger is issued, so time and energy binning is limited by counting statistics only. The on-board trigger logic is active on several time window spanning from 300~$\mathrm{\mu}s$ to 8.192~s and is described in details in \cite{Fuschino2008}.

It must be noted that among the several GRBs localized by SuperAGILE, only seven of them have been also detected by MCAL, mainly because of the different sensitivity in different energy bands and the different angular response of the two instruments. The MCAL in-flight performance and first GRB detections are described in \cite{Marisaldi2008,Marisaldi2008b}.

\subsection{GRB detection and search for Upper Limits with the GRID}

After two years of operations three GRBs were detected by AGILE-GRID: 080514B \cite{Giuliani2008},
090401B \cite{Moretti2009} and 090510 \cite{Longo2009}. A rough estimate of 1 GRB/month that AGILE could have detected in one year of operation was based on the previous EGRET detections \cite{Longo2007}. Actually, GRBs in the MeV--GeV range appear to be quite a rare phenomena, tipically associated to the brightest events, as also confirmed by the recent $Fermi$ detections. 

\begin{figure}[!t]
  \centering
  \includegraphics[width=3.in]{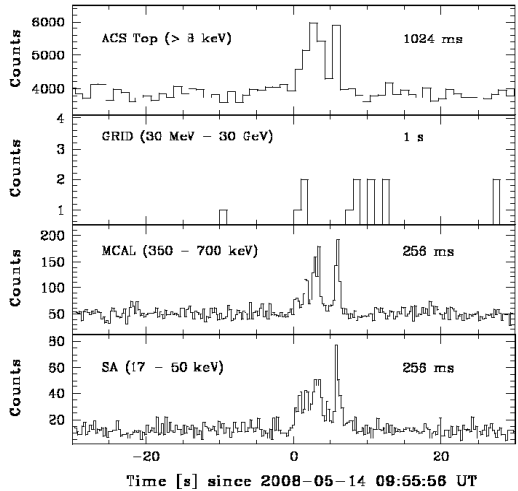}
  \caption{Light curves of GRB~080514B, for the different instruments on-board AGILE. }
   \label{fig1}
\end{figure}

GRB~080514B is the first GRB detected with a pair-conversion tracker telescope above several tens of MeV more than ten years after the EGRET era. Its localization has been a joint effort of the AGILE team and the IPN, since it was only in the 1D-coding field of view of SuperAGILE. Prompt $Swift$ follow-up observations allowed to detect the X-ray afterglow and, using ground-based photometric optical/NIR and millimeter data from several observatories, a photometric redshift of 1.8 has been estimated \cite{Rossi2008}.
The highest photon energy detected is about 300~MeV, while most of the detected GRB photons have energies in the range 25 -- 50~MeV, consistent with a spectrum with a power law photon index of 2.5. This spectrum is also consistent with the extrapolation of that obtained in the MeV range by $Konus$-WIND and MCAL. This bright burst was detected by all the detectors on-board AGILE, including the anti-coincidence system, as shown by the multiple light curves reported in Figure~\ref{fig1}. 
The most remarkable feature of GRB~080514B is its high energy extended emission, i.e. the fact that the arrival times of the high energy photons detected with the GRID are not coincident with the brightest peak detected in hard X-rays. This feature, potentially capable to place significant constraints on emission models, was already suggested concerning the EGRET bursts, then further confirmed by the Fermi detection of GRB~080916C \cite{grb080916c} and by the AGILE and $Fermi$ detection of GRB~090510. In particular, this latest burst, whose duration as estimated by MCAL is about 200~ms, is the first short GRB detected above 100~MeV by AGILE.

For the remaining several tens of events in the AGILE field of view localized and detected by other satellites ($Swift$, INTEGRAL, {\it Fermi} GBM, the IPN) or by SuperAGILE a study of the upper limits in the high energy band is under consideration. 
%From the preliminary estimates, a typical the upper limit in count fluence is around 0.01 - 0.04 ph cm$^-2$. 

\section{Very short time scales detection capabilities}

The trigger logic for burst search on time scales shorter than 64~ms (16~ms, 1~ms and sub-millisecond) is managed by a dedicated hardware section of the PDHU because the fast timing requirements could not be met by a software task. 
For these time scales neither spatial nor energy segmentation is applied for triggering purposes. 
The sub-millisecond time window has an implementation different from that of the other time windows, and from a typical ratemeter time binning in general. In fact, for every detected event a time window is opened (with duration between 300~$\mathrm{\mu}s$ and 1~ms, selectable by tele-commands) and a trigger is issued if a number of events above a fixed threshold is detected in that time window. The novelty of this approach is that, since a new search time window is opened starting from every event, no trigger can be lost due to a mismatch with the time binning. This is particularly important for the search for faint but very timely localized bursts of events.
The trigger logic on the 16~ms and sub-millisecond time windows has been enabled and configured since 20 June 2008. The trigger logic on the 1~ms time window was steadily configured on 2 March 2009, due to the requirement for a software upgrade. With the current settings, a fixed threshold of 23 events is set for the 16~ms time window, the sub-millisecond time window is set to  300~$\mathrm{\mu}s$ with a fixed threshold of 7 events, and a fixed threshold of 8~events is set for the 1~ms time window. With this configuration about 70 triggers/day are obtained. Because of telemetry availability, it has been decided to release the trigger criteria (i.e. keep the threshold as low as possible) in favor of on-ground selection. With few tens of events/trigger it is often challenging to discriminate the nature of these events and disentangle triggers of cosmic, geophysical or instrumental origin, so a careful trigger validation and rejection strategy must be applied.

\begin{figure}[!t]
  \centering
  \includegraphics[width=3.in]{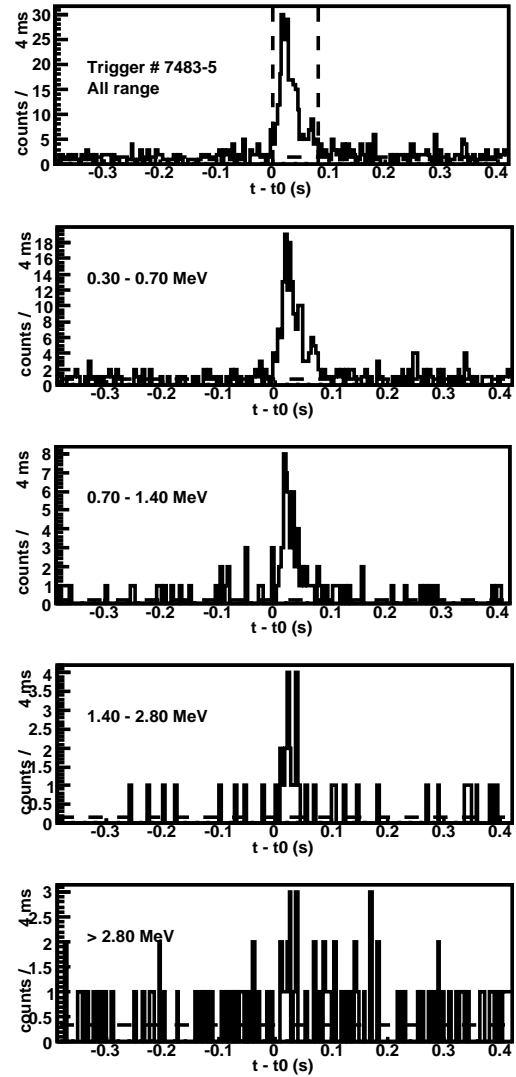}
  \caption{MCAL light curves for GRB~081004, in  different energy bands. }
   \label{fig2}
\end{figure}

By the way several confirmed cosmic GRBs have been detected in the 16~ms time scale. 
Figure \ref{fig2} shows the MCAL light curves for GRB~081004 (not reported in GCN, trigger time $\mathrm{t_0}$ 04 Oct. 2008, 13:26:16 UT), for different energy bands. This burst was triggered on the 16~ms time window, has a double peaked structure, a total duration of about 80~ms and exhibits significant emission above 3~MeV. Its high peak flux, although the limited overall fluence, allowed a time binning of 4~ms.

Due to the combined characteristics of
   \begin{itemize}
\item[-] an energy range extended up to several tens of MeV, 
\item[-] a trigger logic on very short time scales,
\item[-] the photon-by-photon download capabilities, 
  \end{itemize}
MCAL is a suitable instrument to detect also Terrestrial Gamma Flashes (TGFs). TGFs are very hard and bright millisecond-scale bursts of gamma-rays correlated to thunderstorm activity, discovered by BATSE \cite{Fishman1994} and observed also by the RHESSI experiment \cite{Smith2005}. 
A preliminary trigger selection allows to detect about 4~TGF candidates/month, this rate doubled after the onset of the trigger logic on the 1~ms time scale. The characteristics of the selected population is compatible with that provided by RHESSI, and a dedicated publication is in preparation.

\section{Conclusions}

The AGILE satellite is capable to detect GRBs in an energy range extended over 6 orders of magnitude. Since its launch AGILE detected more than 120 GRBs, 21 of them localized by SuperAGILE and 3 of them detected by the GRID above several tens of MeV. 
In particular, the analysis of GRB~080514B, the first GRB detected above 100~MeV since the EGRET era, evidenced a characteristic extended emission in gamma-rays, not synchronized with the peak emission in hard X-rays, which seems to be a common feature of high energy GRBs. Another strength point of AGILE is the MCAL trigger logic implemented on time windows as short as  300~$\mathrm{\mu}s$. To our knowledge it is the first time that a trigger logic on sub-millisecond time scales is implemented in a space instrument. In addition to very short burst of cosmic origin, this peculiar design allowed the detection of terrestrial gamma-ray flashes, thus extending the scientific exploitation of AGILE to the geophysics field.

\section*{Acknowledgments}

The AGILE Mission is funded by the Italian Space Agency (ASI) with scientific and programmatic participation by the
Italian Institute of Astrophysics (INAF) and the Italian Institute of Nuclear Physics (INFN).

\end{document}